\documentclass[prl,aps,twocolumn,floatfix,amsmath,amssymb,groupedaddress]{revtex4-1}

\usepackage{verbatim}
\usepackage{latexsym}
\usepackage{graphicx}
\usepackage{tabularx}
\usepackage{times}
\usepackage[usenames,dvipsnames]{color}
\usepackage{amsmath}
\usepackage{dcolumn}
\usepackage{latexsym,amsmath,amssymb,bm,euscript}
\usepackage[
	colorlinks=true,
	urlcolor=BlueViolet,
 citecolor=Maroon,
	plainpages=false,
 	pdfpagelabels,
 	bookmarksnumbered
 	]{hyperref}

\newcommand{\sgn}{\textrm{sgn}}

\def\beq{\begin{equation}}
\def\eeq{\end{equation}}

\begin{document}

\title{Theory of Disorder-Induced Half-Integer Thermal Hall Conductance}
\author{David F. Mross}
\author{Yuval Oreg}
\author{Ady Stern}
\author{Gilad Margalit}
\author{Moty Heiblum}
\affiliation{Braun Center for Submicron Research, Department of Cond. Matter Physics, Weizmann Institute of Science, Rehovot 76100, Israel}

\begin{abstract}
 Electrons that are confined to a single Landau level in a two dimensional electron gas realize the effects of strong electron-electron repulsion in its purest form. The kinetic energy of individual electrons is completely quenched and all physical properties are dictated solely by many-body effects. A remarkable consequence is the emergence of new quasiparticles with fractional charge and exotic quantum statistics of which the most exciting ones are non-Abelian quasiparticles. A non-integer quantized thermal Hall conductance $\kappa_{xy}$ (in units of temperature times the universal constant $\pi^2 k_B^2 /3 h$; $h$ is the Planck constant and $k_B$ the Boltzmann constant) necessitates the existence of such quasiparticles. It has been predicted, and verified numerically, that such states are realized in the clean half-filled first Landau level of electrons with Coulomb repulsion, with $\kappa_{xy}$ being either $3/2$ or $7/2$. Excitingly, a recent experiment has indeed observed a half-integer value, which was measured, however,  to be $\kappa_{xy}=5/2$. We resolve this contradiction within a picture where smooth disorder results in the formation of mesoscopic puddles with locally $\kappa_{xy}=3/2$ or $7/2$. Interactions between these puddles generate a coherent macroscopic state, which is reflected in an extended plateau with quantized $\kappa_{xy}=5/2$. The topological properties of quasiparticles at large distances are determined by the macroscopic phase, and not by the microscopic puddle where they reside. In principle, the same mechanism might also allow non-Abelian quasiparticles to emerge from a system comprised of microscopic Abelian puddles.
\end{abstract}
\maketitle

The fractional quantum Hall (FQH) state at Landau level filling factor $\nu=\frac{5}{2}$ was the first even-denominator state to be observed experimentally~\cite{FiveHalvesDiscovery}. Soon after its discovery it has been suggested that pairing should have an important role in its formation~\cite{MooreRead,ReadGreen}, and several possible candidates for the paired state have been proposed~\cite{HalperinQH,Greiter92,MooreRead,ReadGreen,AntipfaffianLevin,AntipfaffianLee}. These states differ in several properties, with the most important distinction being between Abelian and non-Abelian states.

While the topological properties of these states have been thoroughly studied theoretically, their identification for the experimentally realized $\nu=\frac{5}{2}$  has proven to be difficult~\cite{SternPerspective}. Numerically, exact diagonalization~\cite{Morf98,HaldaneRezayi00,Peterson08,Jain10,Morf10,RezayiSimon11,Troyer15}
and DMRG studies~\cite{Sarma08,Sarma09} have pointed towards the non-Abelian Pfaffian state, proposed by Moore and Read~\cite{MooreRead,ReadGreen}, and its particle-hole (PH) conjugate state (the ``Anti-Pfaffian" \cite{AntipfaffianLevin,AntipfaffianLee}) as the most likely ground states. 

\begin{figure}[ht]
\includegraphics[width=\columnwidth]{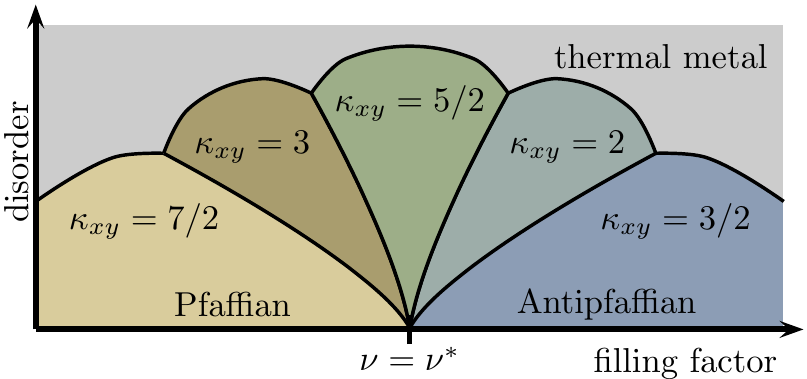}
\caption{{\bf Proposed phase diagram of the approximately half-filled first excited Landau level}. In the absence of disorder, the ground state is either Pfaffian or Anti-Pfaffian, depending on whether the system is more particle-like $(\nu \lesssim \nu^*)$ or hole-like $(\nu \gtrsim \nu^*)$ with an expected first-order transition at $\nu=\nu^* \approx \frac{5}{2}$. In the presence of disorder, this transition splits into four \textit{continuous} phase transitions where the thermal Hall conductance changes by $\Delta \kappa_{xy}=1/2$ (in units of temperature times the universal constant $\pi^2 k_B^2 /3 h$). For even stronger disorder, the system may enter a thermal metal phase, where $\kappa_{xy}$ is non-quantized. In contrast, the electrical Hall conductance $\sigma_{xy}= \frac{5}{2} \frac{e^2}{h}$ remains quantized across these transitions.}
\label{fig:phasedia}
\end{figure}

Several experiments that were carried out in order to differentiate between the candiate states. These include tunneling into the edge~\cite{Radu08,Lin12,Baer14,Fu2016};  noise measurements as a probe for the charge of the elementary excitations~\cite{Dolev2008} (quasiparticles, which were found to be $e^*=\frac{e}{4}$) and for the existence of upstream neutral modes~\cite{Bid2010,UpstreamGross}; interference~\cite{Willett1}; and---most recently---a measurement of the thermal Hall conductance~\cite{moty2017}. Of these, the most unambiguous observation is that of the thermal Hall conductance, which is a topologically protected property whose quantized value differs between the candidate states. It was found experimentally to take the value of $\kappa_{xy}= 5/2$ \cite{moty2017}. The half-integer value of $\kappa_{xy}$ identifies the state as non-Abelian. It is, however, inconsistent with both the Pfaffian and the Anti-Pfaffian states, for which the values of $7/2$ and $3/2$ are expected. Rather, the value of $5/2$ is consistent with a non-Abelian state coined as the particle-hole Pfaffian (PH-Pfaffian), also known as T-Pfaffian \cite{ChenTO} in the context of topological insulator surfaces. This phase first appeared in Ref.~\onlinecite{AntipfaffianLee}, and was recently reinterpreted in the context of Dirac composite fermions \cite{Son}. A prototypical wave function was proposed in Ref.~\cite{zucker}, which also argued PH-Pfaffian to be the most likely phase in light of earlier experiments. There, the ($113$) state \cite{Feldman2014} with $\kappa_{xy}=2$ was also suggested as a possibility, but deemed less likely.

In this work we propose a resolution to the conundrum posed by the discrepancy between the numerical predictions and the experimental results. We focus on an ingredient that is undeniably present in experimental systems, but absent in numerical studies: quenched disorder. The possible stabilization of the PH-Pfaffian phase by disorder has been introduced in Ref.~\onlinecite{zucker}. Our picture starts from the observation that when PH symmetry is present, i.e., in the absence of Landau-level mixing, the Pfaffian and Anti-Pfaffian, which are related by PH transformation, are degenerate at $\nu=\frac{5}{2}$. We expect that even in the absence of an exact PH-symmetry, they become degenerate at $\nu^*\approx \frac{5}{2}$. Upon deviating from this filling, i.e., for $\nu=\nu^* + \delta \nu$, we expect that $\delta\nu<0$ favors Pfaffian while $\delta\nu >0$ favors Anti-Pfaffian.
As a function of $\nu$, the numerical results \cite{Morf98,HaldaneRezayi00,Peterson08,Sarma08,Sarma09,Jain10,Morf10,RezayiSimon11,Troyer15} indicate  that in a clean system there is a direct transition between the two. Since this transition results
from the two competing states being degenerate in energy, it is likely to be of first order.

In the presence of smooth density variations, we then expect puddles of the Pfaffian and the Anti-Pfaffian to form.
The size of these puddles is much larger than the magnetic length, but smaller than the sample size. The electronic Hall conductance is $\sigma_{xy}=\frac{5}{2}\frac{e^2}{h}$ independent of the relative areas occupied by the two phases, since both phases have the same $\sigma_{xy}$. In contrast, the thermal Hall conductance is determined by the predominant phase to be either $\kappa_{xy}^\text{Pfaffian}=7/2  $ or $\kappa_{xy}^\text{Antipf.}= 3/2$.
If the transition between the two cases was direct, $\kappa_{xy}$ would thus jump by 2, as in the clean case. 

Our main result, which is summarized in the schematic phase diagram of Fig.~\ref{fig:phasedia}, is that disorder splits the direct transition into a sequence of four transitions, each characterized by a change of $1/2$ in the thermal Hall conductance. Between the transitions, $\kappa_{xy}$ is sharply quantized to the values $3/2,2,5/2,3$, and $7/2$.  For weak disorder, the electronic density, as expressed by the filling factor $\nu$, functions as a tuning parameter that drives the system across the four continuous quantum phase transitions. For fixed $\nu$ but increasing  strength of disorder, the system may eventually transition into a thermal metal phase where the thermal Hall conductance is no longer quantized \cite{ChoFisher,SenthilFisher2000,Bocquet,ReadLudwig,ChalkerRead}.

\begin{figure}[ht]
\includegraphics[width=\columnwidth]{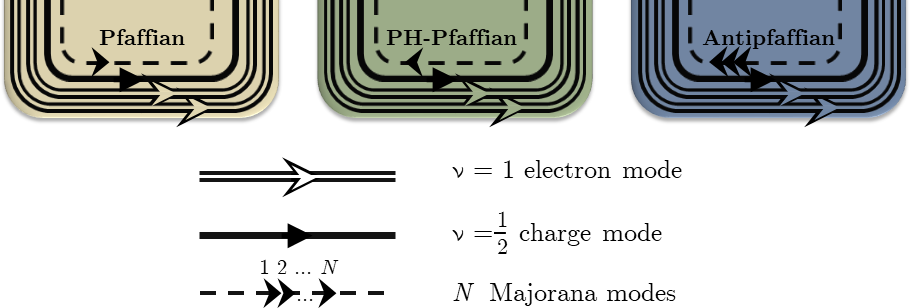}
\caption{{\bf Possible edge modes of three kinds of Pfaffian phases.} All three phases have identical charge modes, but differ in the number and chirality of neutral Majorana modes (indicated by the number and direction of arrows) and consequently in their thermal Hall conductance.}
\label{fig:pfaffians}
\end{figure}

{\bf \emph{Pfaffianology.}}
Five candidate states for the filling of $\nu=\frac{5}{2}$ are relevant to the present work. Each of these hosts multiple edges modes, which interact among themselves. Beyond a length scale $\ell_\text{eq}$ the edge modes reach mutual equilibrium, at which point the state achieves a quantized thermal Hall conductance \cite{Kane1994}. For the purpose of clarity, we will henceforth assume that all relevant length scales exceed $\ell_\text{eq}$. (The values of $\ell_\text{eq}$, as well as other microscopic length scales, depend on non-universal properties and are hard to estimate. Fortunately, these values are not crucial to our main results; we will return to this point after explaining the phase diagram.)

All the states we consider have two chiral ($\nu=1$) electron edge modes of the two filled Landau levels, each contributing $\sigma_{xy}=1$ and carrying central charge $c=1$, as well as one co-propagating ($\nu=\frac{1}{2}$) charge mode contributing $\sigma_{xy}=\frac{1}{2}$ and carrying $c=1$; these modes contribute the required electrical Hall conductance, as well as a thermal Hall conductance of $\kappa_{xy}=3$.
The states differ by the number $n_M$ of neutral Majorana modes (with a negative number indicating upstream modes), which determines the total thermal Hall conductance to be $\kappa_{xy}=3+n_M/2$. The values of $n_M$ for all relevant phases are listed in Table~\ref{tab:phases}, and the edge modes of the non-Abelian phases are depicted in Fig.~\ref{fig:pfaffians}. Notice that only the PH-Pfaffian ($n_M=-1$) is compatible with PH-symmetry under which  $\kappa_{xy}-\kappa_{xy}^{\nu=2}\rightarrow 1- (\kappa_{xy}-\kappa_{xy}^{\nu=2})$, with $\kappa_{xy}^{\nu=2}=2$. The corresponding edge states can be succinctly expressed in terms of their Lagrangian density
$ {\cal L}^\chi_\text{vac.-QH} = {\cal L}^\chi_\text{c}+ {\cal L}^\chi_\text{n}[n_M]$ with
\begin{align}
{\cal L}^\chi_\text{c} = & \frac{1}{4\pi} \sum\nolimits_{ij}\left[\chi K_{ij}\partial_t \phi_i\partial_x \phi_j- V_{ij}(\partial_x \phi_i)(\partial_x \phi_j) \right]~,\nonumber\\
{\cal L}^\chi_\text{n}[n_M]=& \sum\nolimits_{i=1}^{|n_M|} \gamma_i(\partial_t -\chi \sgn(n_M)v \partial_x)\gamma_i~,
\label{eq:Edge_Lag}
\end{align}
where $\phi$ are bosonic modes and $\gamma$ are Majorana fermions; $V$ is positive definite, $K=\text{diag}(1,1,2)$, $v>0$ and $\chi=\pm 1$ (determining the chirality of the edge modes). An overview of the various edge modes appears in Ref.~\onlinecite{Feldman2013} (see also Fig.~6 in the extended material of Ref.~\cite{moty2017}).

\begin{table}[ht]
  \centering
    \caption{{\bf Summary of $\mathbf{\boldsymbol{\nu}\approx 5/2}$ phases relevant for our discussion}. The number $n_M$ of Majorana modes determines the thermal Hall conductance according to $\kappa_{xy} = 3 + n_M/2$.}
  \begin{tabular}{c|ccccc}
  \hline\hline
   QH Phase& Pfaffian& $K=8$ & PH-Pfaffian & (113) & Anti-Pfaffian\\
    \hline\hline
 $n_M$ & 1& 0 & -1 & -2 & -3\\
  \hline
$\kappa_{xy}$& $7/2$& $3$& $5/2$ & $2$ & $3/2$\\
  \hline\hline
  \end{tabular}
  \label{tab:phases}
\end{table}

\begin{figure}[ht]
\includegraphics[width=.47226\columnwidth]{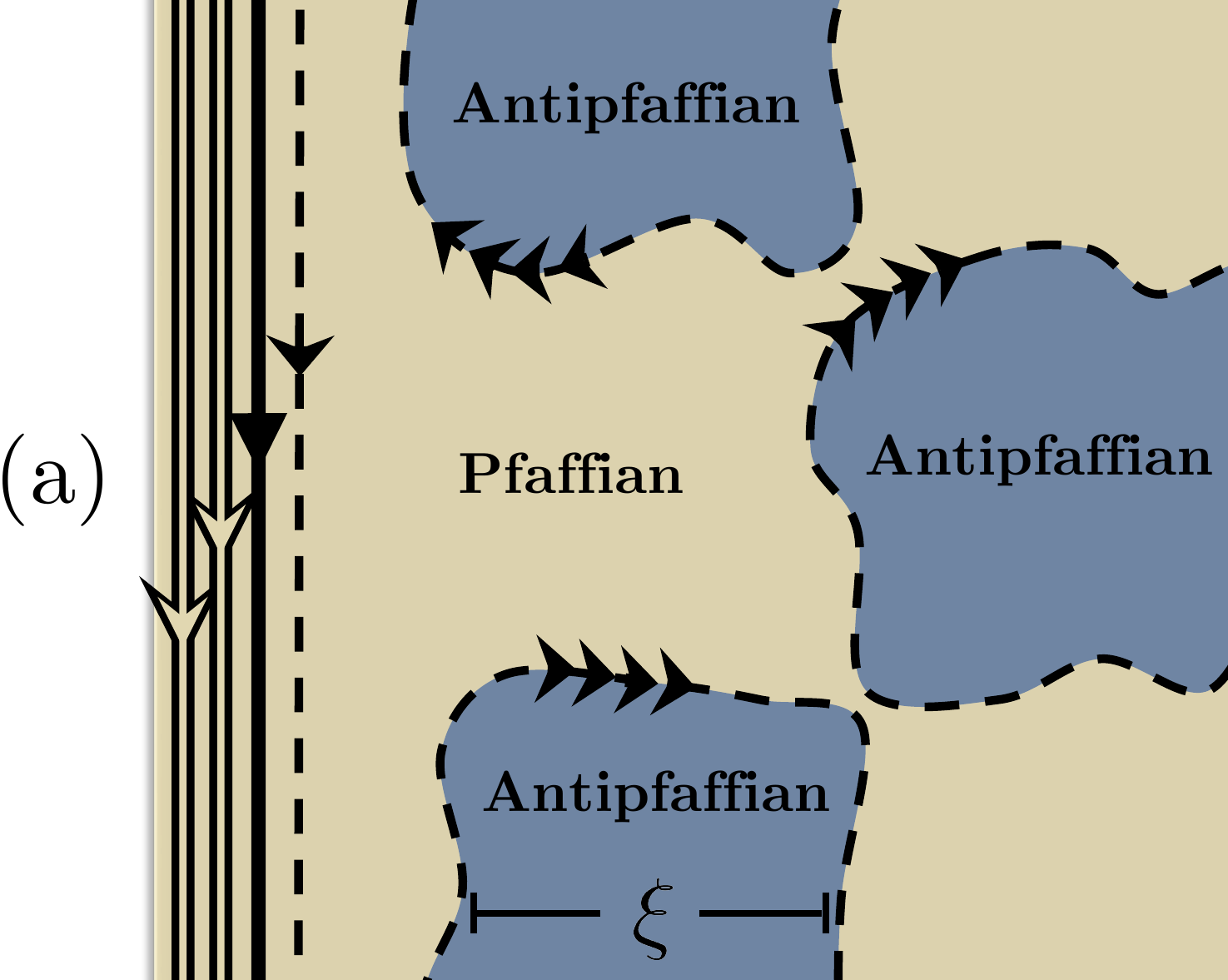}\hspace{4mm}
\includegraphics[width=.447738\columnwidth]{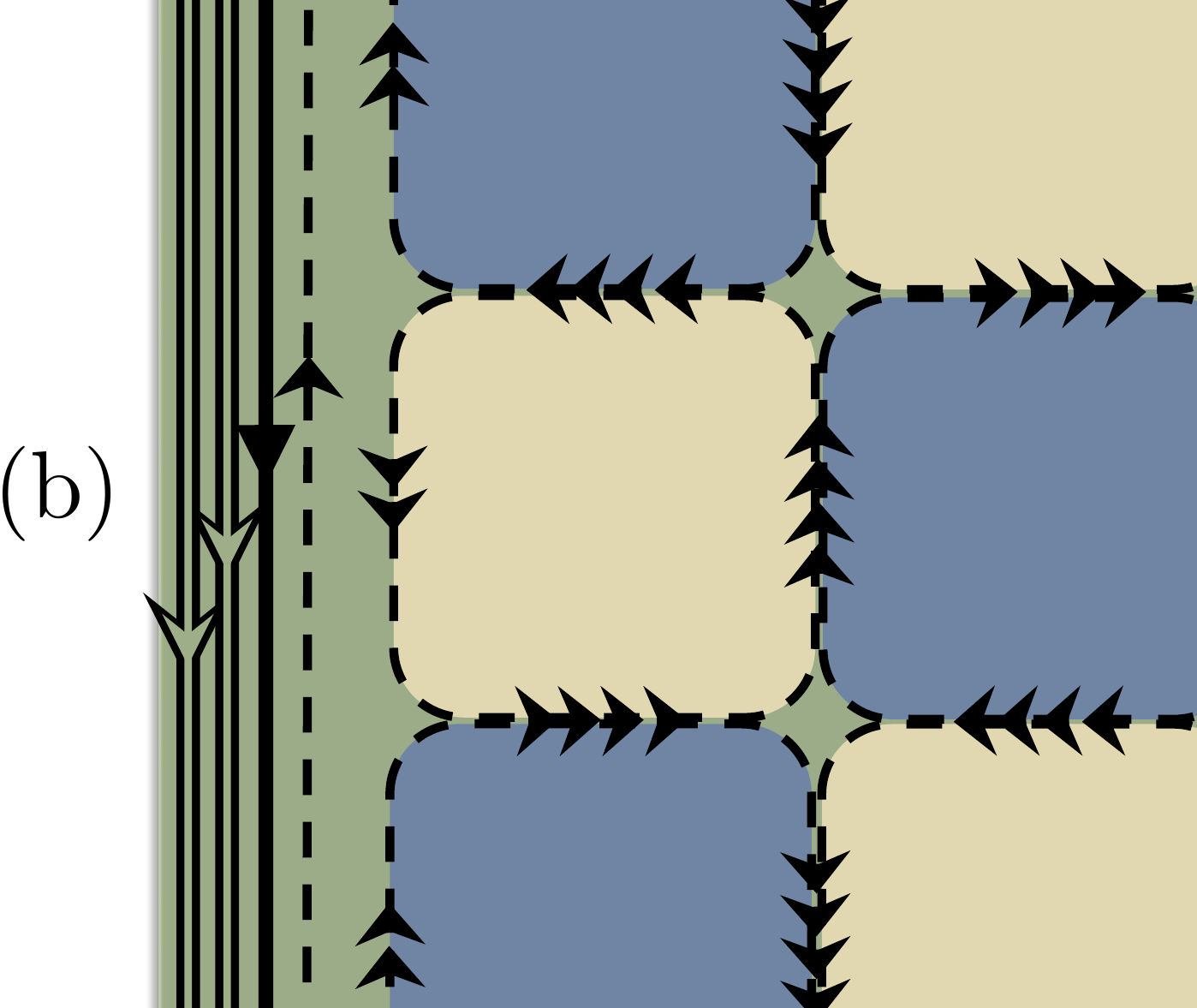}
\caption{{\bf Microscopic puddles and schematic network model.} (a) Puddles of Anti-Pfaffian (of size $\sim\xi$) embedded in a Pfaffian phase. Each Pfaffian--Anti-Pfaffian boundary hosts four co-propagating Majorana fermions.  (b) Adding a topologically trivial pair of counter-propagating Majorana modes next to the sample boundary, followed by suitable hybridization of counter-propagating modes results in the PH-Pfaffian edge on top of a Chalker-Coddington model containing four Majorana fermions.}
\label{fig:network}
\end{figure}

{\bf \emph{Disorder and network model.}}  In our model, we consider a disorder potential that is sufficiently weak and/or smooth that the results of numerical studies apply locally and favor the formation and Pfaffian or Anti-Pfaffian puddles, see Fig.~\ref{fig:network} (a). 

The Pfaffian--Anti-Pfaffian boundary is captured by \cite{MaissamCFL}
\begin{align}
 {\cal L}^\chi_\text{Pf.-APf.} =&{\cal L}^\chi_\text{c}+{\cal L}^\chi_\text{n}[1] +{\cal L}^{-\chi}_\text{c} +{\cal L}^{-\chi}_\text{n}[-3].\label{eqn.pfapf}
\end{align}
Notice that when the two quantum Hall states are separated by vacuum, fractional excitations cannot tunnel between the two. Still, the non-chiral Lagrangian of the charge modes ${\cal L}^\chi_\text{c} +{\cal L}^{-\chi}_\text{c}$ can be gapped by sufficiently strong tunneling of (pairs of) \textit{electrons} across the Pfaffian--Anti-Pfaffian boundary \cite{MaissamCFL}. This `stitches together' the two quantum Hall states and permits fractional excitations to traverse between the two.  However, the neutral Majorana fermions co-propagate and thus cannot be gapped out. Instead, the two neutral terms in Eq.~\eqref{eqn.pfapf} combine into ${\cal L}^\chi_\text{n}[4]$. Consequently, the Pfaffian--Anti-Pfaffian boundary is described by four co-propagating Majorana fermions, which have an $O(4)$ symmetry of rotating between them.

Our proposed model to describe a disordered $\nu=\frac{5}{2}$ system may be viewed as a checkerboard of alternating Pfaffian and Anti-Pfaffian states as shown in Fig.~\ref{fig:network} (b), with random scattering at each vertex. (This is a generalization of the Chalker-Coddington network model, which has been highly successful in describing integer-quantum Hall plateau transitions \cite{Chalker}). The Majorana modes are neutral and subject to short-range interactions only (unlike the case of electrons in the integer quantum Hall plateau transitions that may experience long-range interactions). As we review below, the critical point of a network model with non-random vertices is described by four Majorana cones, where such interactions are irrelevant. We therefore neglect them henceforth.

To gain intuition for the underlying physics, it is instructive to first analyze the strongly anisotropic limit of the network model. This model consists of (narrow) infinite strips that alternate between Pfaffian and Anti-Pfaffian. (A closely related model was studied in Ref.~\cite{MaissamCFL}, which  focused on the composite Fermi liquid.) The low-energy properties of the model are encoded in ${\cal L}^\text{anis} = \sum_y \left[{\cal L}_\text{$y$}^\text{edge} +{\cal L}_\text{$y$}^\text{tunnel} \right]$ with

\begin{align}
{\cal L}_\text{$y$}^\text{edge} &=  \vec \gamma_{y}\cdot(\partial_t - (-1)^yv \partial_x)\vec\gamma_{i,y}~,\label{anis}\\
{\cal L}_\text{$y$}^\text{tunnel}&=i \vec \gamma_{y} \left[M^\text{unif}+(-1)^y M^\text{stag}+M^\text{rand}_{y}(x)\right]\vec \gamma_{y+1}\nonumber,
\end{align}
where $\vec \gamma^T_y=(\gamma_{1,y},\gamma_{2,y},\gamma_{3,y},\gamma_{4,y})$ and $y$ enumerates Pfaffian-Anti-Pfaffian interfaces.  The tunneling terms are parameterized by three $4\times 4$ matrices, describing uniform ($M^\text{unif}$), staggered ($M^\text{stag}$) and random ($M^\text{rand}$) tunneling terms. The $O(4)$ symmetry of the clean case makes $M^\text{unif}$ proportional to the unit matrix.  When $M^\text{rand}=M^\text{stag}=0$, ${\cal L}^\text{anis}$ features a discrete translation symmetry and is readily diagonalized in momentum space where one finds four gapless Majorana cones.

 Perturbing these cones with a generic mass matrix $M^\text{stag}$ opens an energy gap and the resulting phases can be classified according to the number of negative mass terms. A global Anti-Pfaffian phase arises when all four cones are gapped by tunneling across Pfaffian strips (e.g., in the limit where these strips are much narrower than Anti-Pfaffian strips). We adopt the convention that this corresponds to all masses being negative for the Majorana fermions. When $M^\text{stag}$ is $O(4)$ symmetric, there is a direct transition between a Pfaffian and an Anti-Pfaffian. When that is not the case, flipping a sign of a single mass term to be positive increments the total thermal Hall conductance by $\Delta \kappa_{xy}=1/2$, and all phases in Table~\ref{tab:phases} may be realized.

We note that a clean uniform system differs from  ${\cal L}^\text{anis}$ without disorder. The transition between Pfaffian and Anti-Pfaffian is expected to be first order in the former, but continuous in the latter. Consequently, intermediate phases with $\kappa_{xy}=2, {5}/{2}$, and $3$ are more readily accessible in the network model, where they arise upon weakly perturbing the critical point. In the presence of disorder, in which edge states between the puddles necessitate closure of the energy gap, we expect this distinction between the two models to become insignificant, and both to exhibit the same universal behavior.

The key features of the anisotropic network model carry over to the isotropic case. Being free-fermion systems without charge conservation or any other symmetry, they fall into class D in the Altland-Zirnbauer classification \cite{Zirnbauer,AltlandZirnbauer}.  In two dimensions, this class is characterized by a $\mathbb{Z}$ topological invariant, i.e., an integer $n_M$ corresponding to a thermal Hall effect that is quantized as $\kappa_{xy} = n_M/2$~\cite{RyuTenfoldWay,Kitaevperiodictable}. Certain extra symmetries, such as the $O(4)$ symmetry in Eq.~\eqref{anis}, can ensure transitions where $\kappa_{xy}$ jumps by $2$. However, this changes when disorder respects the protecting symmetry only \textit{on average}. Random scattering completely mixes the four species of Majorana fermions and at large enough length scales, the system is effectively comprised of a single species of Majorana fermions. Without fine-tuning, we thus expect that such a system exhibits phase transitions across which the thermal Hall conductance changes by $\Delta \kappa_{xy}=1/2$ (and not by a larger $\Delta \kappa_{xy}$), leading to the proposed phase diagram shown in Fig.~\ref{fig:phasedia}. The behavior around each of the four lines that emanate from the transition point of the clean system was studied in Refs.~\onlinecite{ChoFisher,SenthilFisher2000,Bocquet,ReadLudwig,ChalkerRead}, where the phase diagram of a network model with a single species of Majorana fermions was analyzed.

In addition to the localized phases with quantized $\kappa_{xy}$, a thermal metal phase with non-quantized $\kappa_{xy}$ can also arise \cite{ChoFisher,SenthilFisher2000,Bocquet,ReadLudwig,ChalkerRead}. There are thus two scenarios for a disordered system to transition between $\kappa_{xy}=7/2$ (Pfaffian) and $\kappa_{xy}=3/2$ (Anti-Pfaffian): (i) The Majorana fermions in the bulk remain localized apart from sharp transitions with $\Delta\kappa_{xy}=1/2$ each, or (ii) There is an intermediate delocalized phase and $\kappa_{xy}$ varies continuously. Which case is realized depends on the type and strength of randomness, with an important role being played by vortex disorder. Ascertaining the fate of a particular system requires a detailed microscopic analysis that we do not attempt here. The experimental observation of a quantized thermal Hall conductance supports the first scenario over the second although it does not exclude the possibility of a metallic bulk with longitudinal thermal conductance $\kappa_{xx} \ll \kappa_{xy}$.

The macroscopic localized phases that arise from a checkerboard of Pfaffian and Anti-Pfaffian puddles can be inferred by considering the sample edge [cf.~Fig.~\ref{fig:network} (a)]. It is instructive to separate the system into two part: An `outer' system that describes all properties that do not change on  the scale of the typical puddle size, $\xi$, and an `inner' system whose properties alternate on this scale. For example, to analyze the formation of a $\kappa_{xy}=5/2$ edge, it is expedient to introduce two pairs of counter-propagating chiral Majorana modes next to the sample boundary. (Such modes could gap one another and thus do not affect the two-dimensional phase.) By appropriately hybridizing counter-propagating modes, one readily achieves the desired separation between uniform and rapidly varying degrees of freedom [see~Fig.~\ref{fig:network} (b)]. The outer system contains all the charge modes corresponding to $\sigma_{xy}=\frac{5}{2}\frac{e^2}{h}$, and an oppositely propagating chiral Majorana mode. The inner system is a checkerboard lattice, where each plaquette is fully encircled by two chiral Majorana modes. When the inner system is localized and symmetric, the sample edge is described solely by the outer system, i.e., its edge modes are the ones of PH-Pfaffian.

The emergent splitting of symmetry-protected phase transitions due to statistically symmetric disorder is familiar from other contexts. One example is the integer quantum Hall plateau transition between $\nu=0$ and $\nu=2$ of spin-degenerate electrons with spin-orbit scattering \cite{LeeChalker}. The introduction of spin-orbit scattering splits the single transition across which the Hall conductance changes by $\Delta \sigma_{xy}=2 \frac{e^2}{h}$ into two transitions, each with  $\Delta \sigma_{xy}= \frac{e^2}{h}$. A second example is that of valley degenerate electrons in graphene, where random inter-valley scattering splits the transition in a similar fashion \cite{Ostrovsky2008}.

To corroborate our scenario, we numerically studied disorder-induced splitting of transitions between different topological phases in two superconducting model systems: a one-dimensional superconductor in class BDI and a two-dimensional superconductor in class D \cite{RyuTenfoldWay,Kitaevperiodictable}.  In the one-dimensional case, we studied four identical Kitaev-chains that undergo phase transitions between having and not having Majorana zero modes at their ends. The $O(4)$ symmetry inherent in this model ensures that there is a single transition, where the topological invariant changes by four. Upon introducing disorder that preserves this $O(4)$ symmetry only on average, we observe numerically that the transition splits into a sequence of four transitions. A two-dimensional system where symmetry ensures a fourfold transition is a bilayer $p_x+ i p_y$ superconductor transitioning into a $p_x- i p_y$ bilayer. This transition involves $\Delta n_M=2$ in each layer, so that in total $\Delta n_M=4$. The transition is protected by a combination of layer-interchange and spatial rotation symmetries. Again, we introduce statistically symmetric disorder and numerically observe splitting of the transition into a sequence of four transitions with $\Delta n_M=1$ (See Suppl. Material for details of both calculations).

We note that our model is based on the picture of a network that is fully coherent. An \textit{ incoherent} mixture of puddles with different electrical  Hall conductances leads to the so called semicircle law~\cite{semicirclelaw}. According to this law, a system composed of two different phases shows, when the population of the phases is close to being equal, a large longitudinal conductance that is of the order of the difference between the Hall conductances of the two phases. We expect an analogous analysis for thermal transport in an incoherent mixture to lead to a similar law. This result is inconsistent with the observed plateau with  $\kappa_{xy} =5/2$. Yet, it is possible that it would describes samples of sizes larger than the coherence length.

We conclude our analysis of the network model by returning to the question of equilibration of edge modes. Above, we always assumed that all edge modes are fully equilibrated at the length scale $\xi$. When $\xi < \ell_\text{eq.}$, the same model can still be used provided that puddles are interpreted as more coarse-grained objects: at a scale larger than $\ell_\text{eq.}$ and $\xi$, one may define the state of a `puddle' of this size to be whichever state is present in the majority.

{\bf \emph{Transmutation from Abelian to non-Abelian statistics.}} A somewhat surprising outcome of our model is that a system made of puddles of two non-Abelian phases may form a macroscopic Abelian phase, and vice versa. Abelian and non-Abelian phases differ in their ground state degeneracy in the presence of quasiparticles and in the braiding properties of their quasiparticles. It is then interesting to examine how these two characteristics  change in the transition from a single puddle to a macroscopic phase.

On short length scales the characteristics of a quasiparticle reflect the state of the microscopic puddle where it resides. However, the topological properties of the macroscopic state can be inferred only when considering quasiparticles whose separation significally exceeds both the puddle size and the localization length of low-energy excitations. Macroscopic degeneracy and non-Abelian braiding would then exist when each bulk quasiparticle carries a localized zero energy Majorana mode. In all the phases that we consider, quasiparticles may be viewed as vortices in class-D superconductors, each harboring $n_M\text{ mod } 2$ Majorana zero modes in their cores. The microscopic value of the number (determined by the puddle hosting the vortex) need not match the one of the macroscopic state. In that case, the difference must be made up by the effect of the vortex on the network of coupled Majorana modes.

\begin{figure}[ht]
\includegraphics[width=\columnwidth]{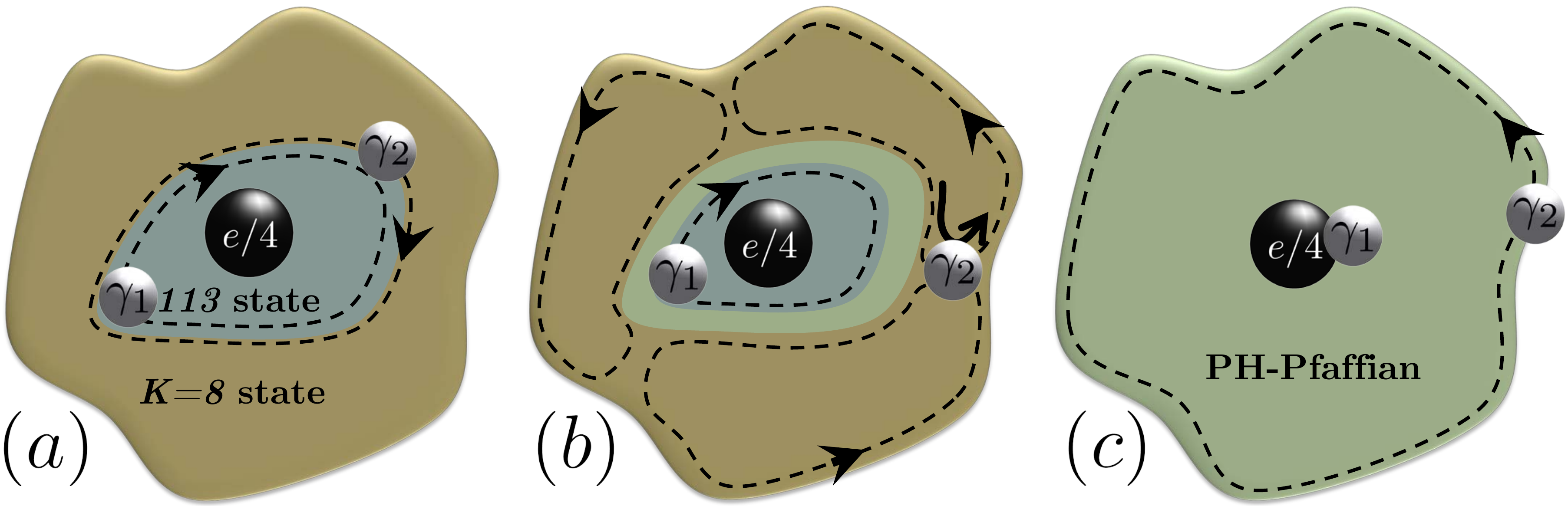}
\caption{{\bf Emergence of non-Abelian quasiparticles from puddles of two Abelian phases.} (a) In the Abelian `host' system, a fractional charge $e/4$ is associated with a pair of Majorana zero modes, which can hybridize with one another. (b) At the transition into the non-Abelian phase, only one of the two Majorana fermions (along with its zero-mode) delocalizes throughout the system. (c) In the non-Abelian phase, the previously percolating Majorana fermion forms an edge state at the outer sample boundary, while still carrying the exact zero-energy mode. }
\label{fig:qp2}
\end{figure}

We illustrate the mechanism behind this in Fig.~\ref{fig:qp2}, starting with puddles of two Abelian phases that form a macroscopic non-Abelian phase.  Here, an $e/4$ excitation changes the boundary conditions of the edge states to create a \textit{pair} of Majorana zero modes at the puddle boundary [cf.~Fig.~\ref{fig:qp2} (a)]. This pair of zero-modes is not protected and could in principle hybridize and gap out. However, when the microscopic puddles of the two Abelian phases generate a macroscopic non-Abelian phase, one of the zero modes must be transfered to the sample boundary, leaving behind a single stable Majorana zero mode bound to the $e/4$ excitation [cf.~Fig.~\ref{fig:qp2} (b) and (c)]. The complementary case of an Abelian phase arising from puddles of two non-Abelian phases can be understood analogously; see the supplementary material for further details.

{\bf \emph{Conclusions and outlook.}} To conclude, there are several competing states that may be adiabatically connected to the ground state of the fractional quantum Hall effect at filling factor $\nu=\frac{5}{2}$ with Hall conductance $\sigma_{xy}=\frac{5}{2} \frac{e^2}{h}$. Theoretically, the states differ by the number of Majorana modes $n_M$ on the edge of the sample, which is reflected in their thermal Hall conductance $\kappa_{xy}=3+n_M/2$.  Numerical works suggest that $n_M=1$ (the Pfaffian) or $n_M=-3$ (the anti Pfaffian) have the lowest energy in \emph{clean} systems. However, previous experiments \cite{Radu08,Lin12,Baer14,Dolev2008,Bid2010,Willett1,Fu2016}, and in particular the recent observation of the normalized thermal Hall conductance of $5/2$~\cite{moty2017} indicate that $n_M=-1$ (the P-H Pfaffian) is the phase observed in the system that was probed experimentally.

In this manuscript we developed a model for the disordered system built from mesoscopic puddles of Pfaffian and Anti-Pfaffian phases. We show, using both numerical and analytical arguments, that a plateau with $\kappa_{xy}=5/2$ is stabilized at sufficiently long distances. Our theory predicts that for moderate  disorder a series of phase transitions between $\kappa_{xy}$ values of $7/2 \rightarrow 3 \rightarrow 5/2 \rightarrow 2 \rightarrow 3/2$ occurs with increasing filling factor $\nu$, as long as the Hall conductance plateau at $\frac{5}{2}\frac{e^2}{h}$ is wide enough. We argue that the properties of the quasiparticles at large distances, and in particular whether their exchange follows Abelian or non-Abelian statistics, are determined by the macroscopic phase, and not by the microscopic puddle in which they reside.

Experimental realization of the full series of transitions that we obtain depends on the width of the $\sigma_{xy}=\frac{5}{2} \frac{e^2}{h}$ plateau, which is usually rather narrow. Both the width of that plateau and the width of each of the phases in Fig. (\ref{fig:phasedia}) increase with disorder (at least for weak disorder), but the scaling of their relative sizes is unknown to us. However, for a mesoscopic system, there may be an alternative route towards observing different quantization of the thermal Hall conductance. The splitting of plateaus is only expected to occur beyond a crossover length scale, and smaller systems exhibit instead the thermal Hall conductance of Pfaffian or Anti-Pfaffian. (A trivial example of this is a system of linear size $\xi$ which contains only a single puddle.) A systematic study of the thermal Hall conductance as a function of sample size could thus be used to test our theory, as well as determine the crossover length scale.

\begin{acknowledgments}
{\bf \emph{Acknowledgments.}} Two upcoming works by D.~Feldman and by B.~Halperin \textit{et al.} study disorder-based mechanisms to explain the experimentally observed $\kappa_{xy}$. We thank them for sharing their insights on this topic. We would also like to thank Jason Alicea, Olexei Motrunich, Pavel Ostrovsky, Eran Sagi, Michael Zaletel and especially Mitali Banerjee for fruitful discussions. This work was performed in part at the Aspen Center for Physics, which is supported by National Science Foundation grant PHY-1607611 (DFM and YO) and partially supported by a grant from the Simons Foundation (DFM); the Israel Science Foundation (DFM, YO, AS, MH); the Minerva foundation with funding from the Federal German Ministry for Education and Research (DFM and MH); the Binational Science Foundation (DFM and YO); the European Research Council under the European Community’s Seventh Framework Program (FP7/2007-2013)/ERC Grant agreement No. 339070 (MH) and Project MUNATOP (YO and AS); Microsoft Station Q (AS); the DFG (CRC/Transregio 183, EI 519/7-1) (YO and AS); the German Israeli Foundation, grant no. I-1241-303.10/2014 (MH).

\end{acknowledgments}

\bibliography{disph}
\newpage

\begin{figure}[ht]
\includegraphics[width=\columnwidth]{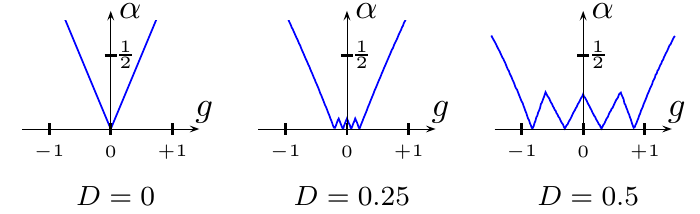}
\caption{{\bf Disorder induced splitting of topological phase transitions in a BDI chain.} Inverse localization length $\alpha$ of a one-dimensional chain in the BDI symmetry class with random hopping terms of variance $D$ that preserve an average $O(4)$ symmetry. In the clean case $D=0$, as a function of the tuning parameter $g$, there is a single phase transition at $g_c=0$ where $\alpha$ vanishes, i.e., where the system is delocalized. For $D\neq 0$ this transition splits into a sequence of four transitions with $g_c(D) \sim D^2$ (see Fig.~\ref{fig:1dloglog}).}
\label{fig:1dmodel}
\end{figure}

{\bf \emph{Disorder-induced splitting of transitions in superconducting model systems.}} We studied numerically two model systems in which disorder splits topological phase transitions in superconductors. Our first model is a one-dimensional superconductor in the BDI class. It is characterized by a topological index $\nu \in \mathbb{Z}$ like class D in two-dimensions and it similarly hosts Majorana fermions at the boundary. Specifically, we consider $H = \sum_n \left[i t_0 \vec \eta_n \cdot \vec \xi_n +i t_1 \vec \eta_n \cdot \vec \xi_{n+1} \right]$ with $N$-component Majorana fermions $\vec \eta$ and $\vec \xi$ on each site. The $O(N)$-symmetry of $H$ ensures a single transition with $\Delta \nu = N$ upon tuning $g=\log (t_0/t_1)$. Disorder is included via $H_\text{dis}= i \vec \eta_n V_{0,n}\cdot \vec \xi_n+i \vec \eta_n V_{1,n}\cdot \vec \xi_{n+1}$ where $V_{0,1}$ are real matrices whose elements are independently Gaussian distributed with zero mean and variance $D$. The arguments provided above predict splitting of the $N$ fold transition. We confirmed this numerically for $N=4$ by computing the smallest Lyapunov exponent (inverse localization length), finding the expected sequence of four transitions with $\Delta \nu=1$ (see Fig.~\ref{fig:1dmodel}).

\begin{figure}[ht]
\includegraphics[width=.7\columnwidth]{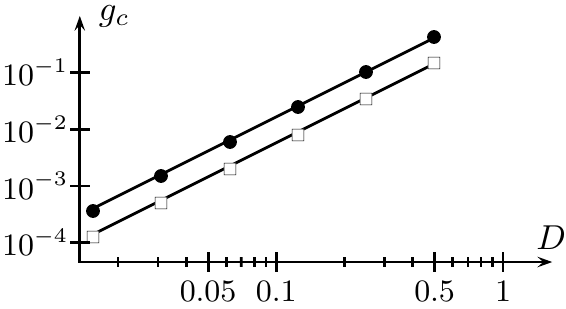}
\caption{{\bf Scaling of the critical coupling in BDI chain with disorder strength.} Critical coupling $g_c$ as a function of disorder $D$. Open boxes and solid squares refer to the two transition seen in Fig.~\ref{fig:1dmodel} for either sign of $g$; both satisfy $g_c \sim D^2$ (solid lines). }
\label{fig:1dloglog}
\end{figure}

\begin{figure}[ht]
\includegraphics[width=\columnwidth]{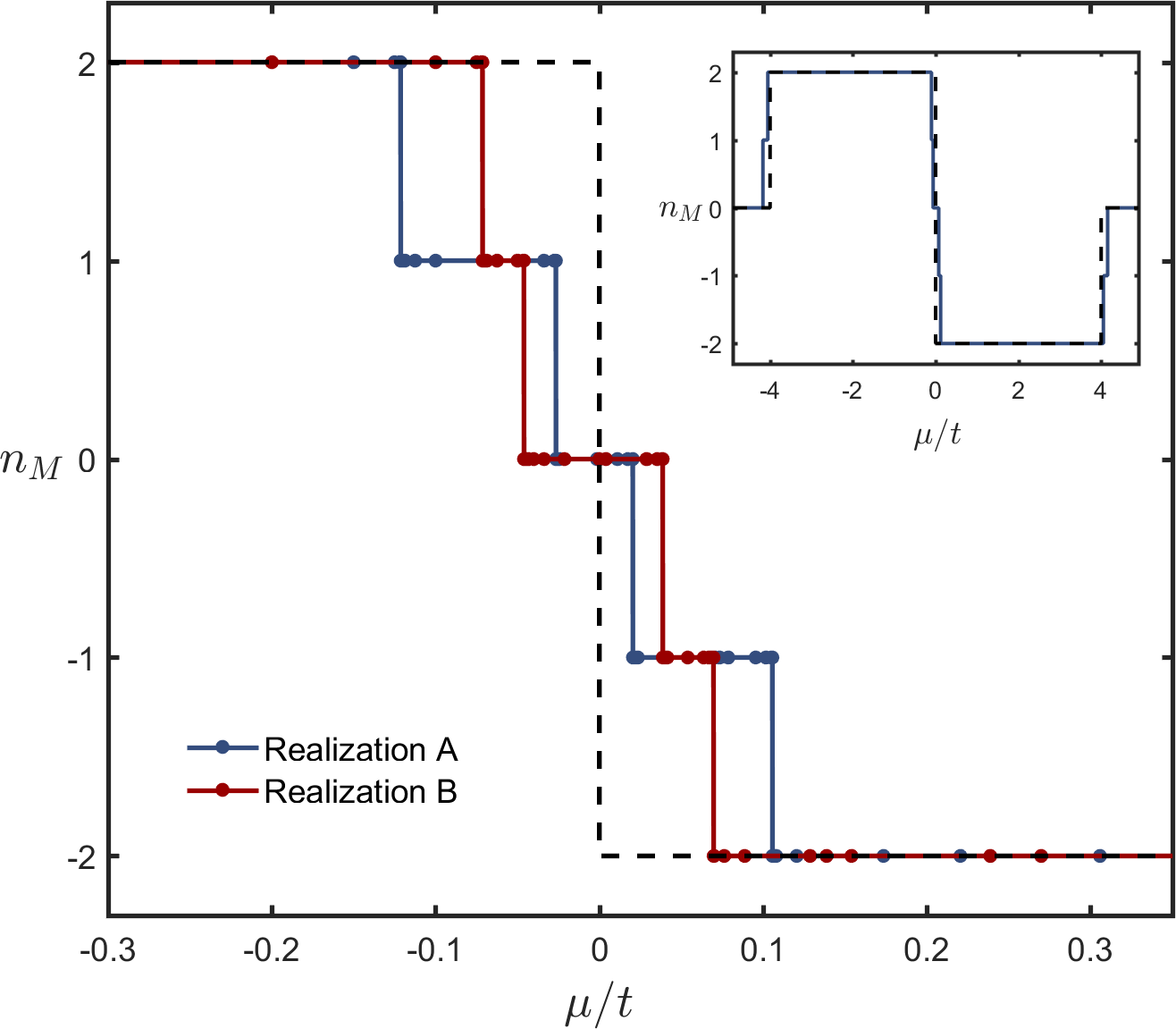}
\caption{{\bf Splitting of topological phase transition in class D due to disorder.} The main figure focuses on the transition near $\mu=0$ for the clean case (dashed line) as well as two disorder realizations (`A' and `B'). The inset shows a wider range of $\mu$ which includes additional phase transitions near $|\mu|=4t$ into topologically trivial $n_M=0$ phases.}
\label{fig:Chern2D}
\end{figure}

Our second example is a two-dimensional superconductor in  class D. Specifically, we consider two layers, $a$ and $b$, of a $p_x+i p_y$ topological superconductor. In the clean case we consider a simple tight-binding Bogoliubov De-Gennes Hamiltonian $H = \frac{1}{2}\sum_{\vec p} \Psi^\dagger_{\vec p} {\cal H}_{\vec p}  \Psi_{\vec p} \nonumber$ with
\begin{align}
\label{eq:2DHamiltonian}
{\cal H}_{\vec p}=&\left[\left(-\mu-2 t \cos p_x -2 t \cos p_y  \right) \tau_z \right.\\
\nonumber&+\left. \Delta\sin p_x \tau_x +\Delta\sin p_y \tau_y \right]  \sigma_0~.
\end{align}
Here $\Psi_{\vec p}^T=\left(a_{\vec p},a^\dagger_{-{\vec p}},b_{\vec p},b^\dagger_{-\vec p}\right)$, $\tau_{\lambda}$ and $\sigma_{\lambda}$ are Pauli matrices for $\lambda=x,y,z$ and the identity for $\lambda=0$. This system features two Majorana cones (one per layer)  at $(p_x,p_y,\mu)$ = $(0,0,-4t),(\pi,0,0),(0,\pi,0)$ and $(\pi,\pi,4t)$. For $|\mu|>4t$ both layers form (trivial) strong-pairing p-wave superconductors without Majorana modes on the edge. For $-4t <\mu< 0$ each layer hosts one chiral Majorana mode on the edge, so $n_M=2$. The chiralities are reversed for $ 0<\mu<4t$, i.e., $n_M=-2$. At $\mu=0$ there are consequently four gapless Majorana cones that characterize a single transition with $\Delta n_M=4$ (this is ensured by a combination of layer-interchange and discrete rotation symmetries). The clean phase is shown in the inset to Fig.~\ref{fig:Chern2D}.

At $\mu=0$, similar to the transition between the Pfaffian and the Anti-Pfaffian states, the Chern number---and therefore the number of Majorana modes---jumps by $4$. We expect that disorder that violates the layer-interchange and spatial symmetries (while preserving them on average) splits the single transition into a sequence of four transitions with $\Delta n_M=1$.

To check this hypothesis we perturb the clean tight-binding Hamiltonian with random on-site terms with matrix structure $\tau_z\sigma_\lambda$ (with $\lambda=0,x,y,z$) and nearest-neighbor terms with matrix structure $\tau_z\sigma_\lambda$ and $i \tau_0 \sigma_\lambda$ , i.e., random hopping but not pairing. We take $t=2 \Delta$ to be real and all the random terms to be uniformly distributed in the interval $[-.4t,.4t]$ (to minimize finite-size effects, we subtract the averages of the random terms). For this model we have calculated the Chern number numerically using the coupling-matrix approach~\cite{CouplingMatrixChernZhang} for systems up to size $16\times16$ . As shown in Fig.~\ref{fig:Chern2D}, we find that the transition from $n_M=2$ to $n_M=-2$ indeed splits. The precise transition points depend on the specific disorder realization in our finite size system, but the splitting into distinct plateaus is generic.

{\bf \emph{Emergence of non-Abelian quasiparticles from Abelian puddles.}}
\begin{figure*}[ht]
\includegraphics[width=.9\columnwidth]{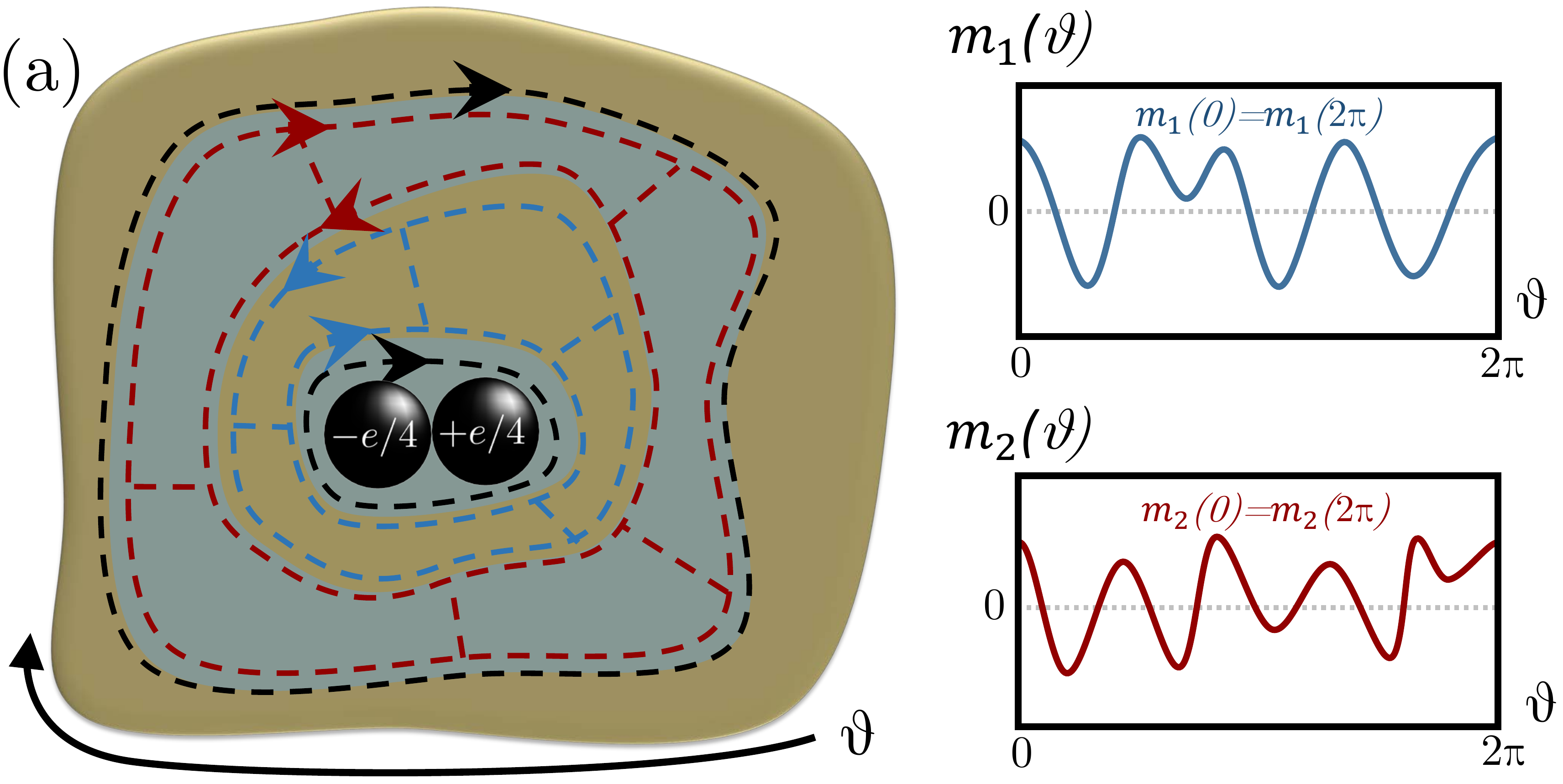}\hspace{8mm}
\includegraphics[width=.9\columnwidth]{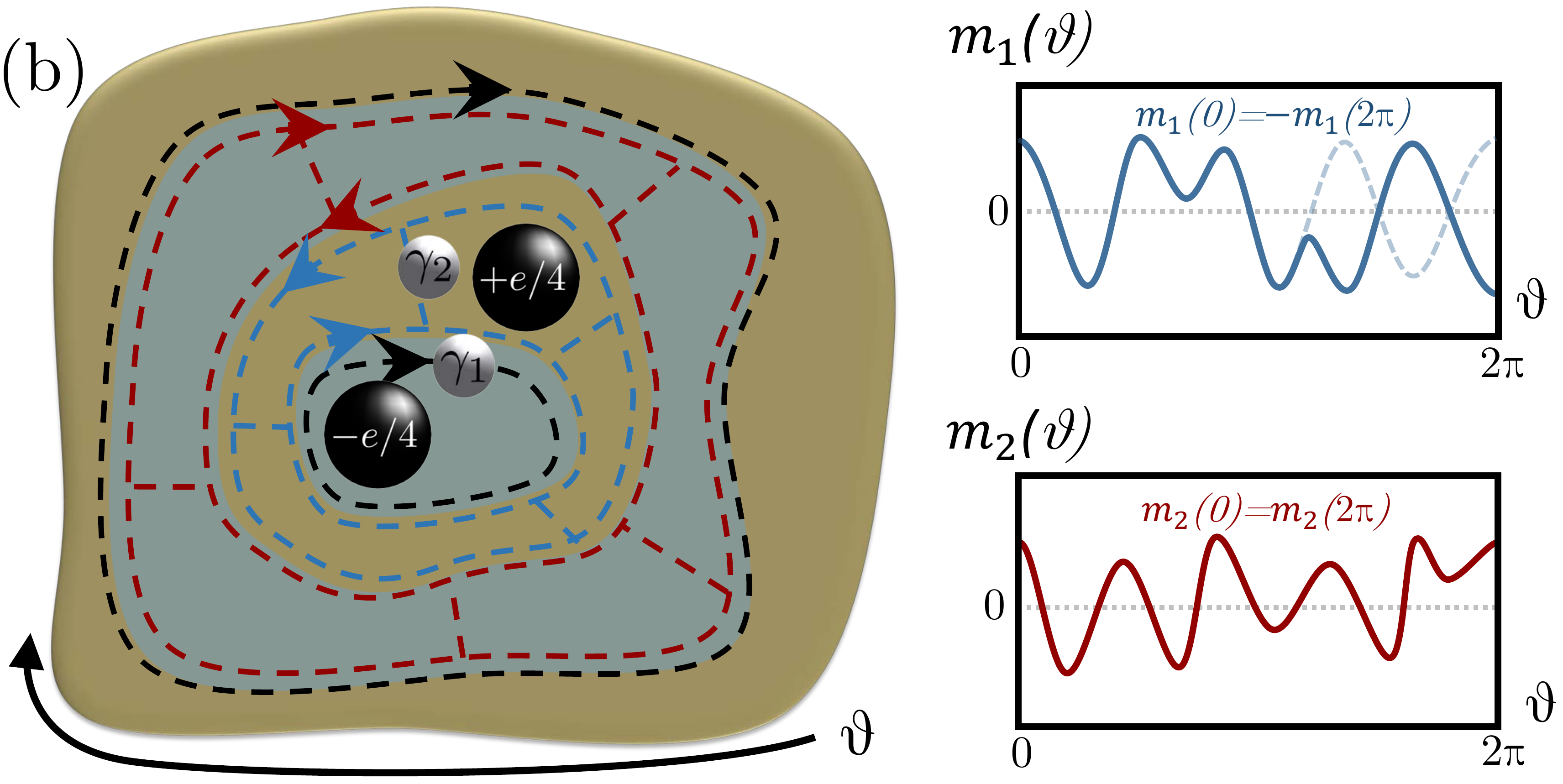}
\includegraphics[width=.9\columnwidth]{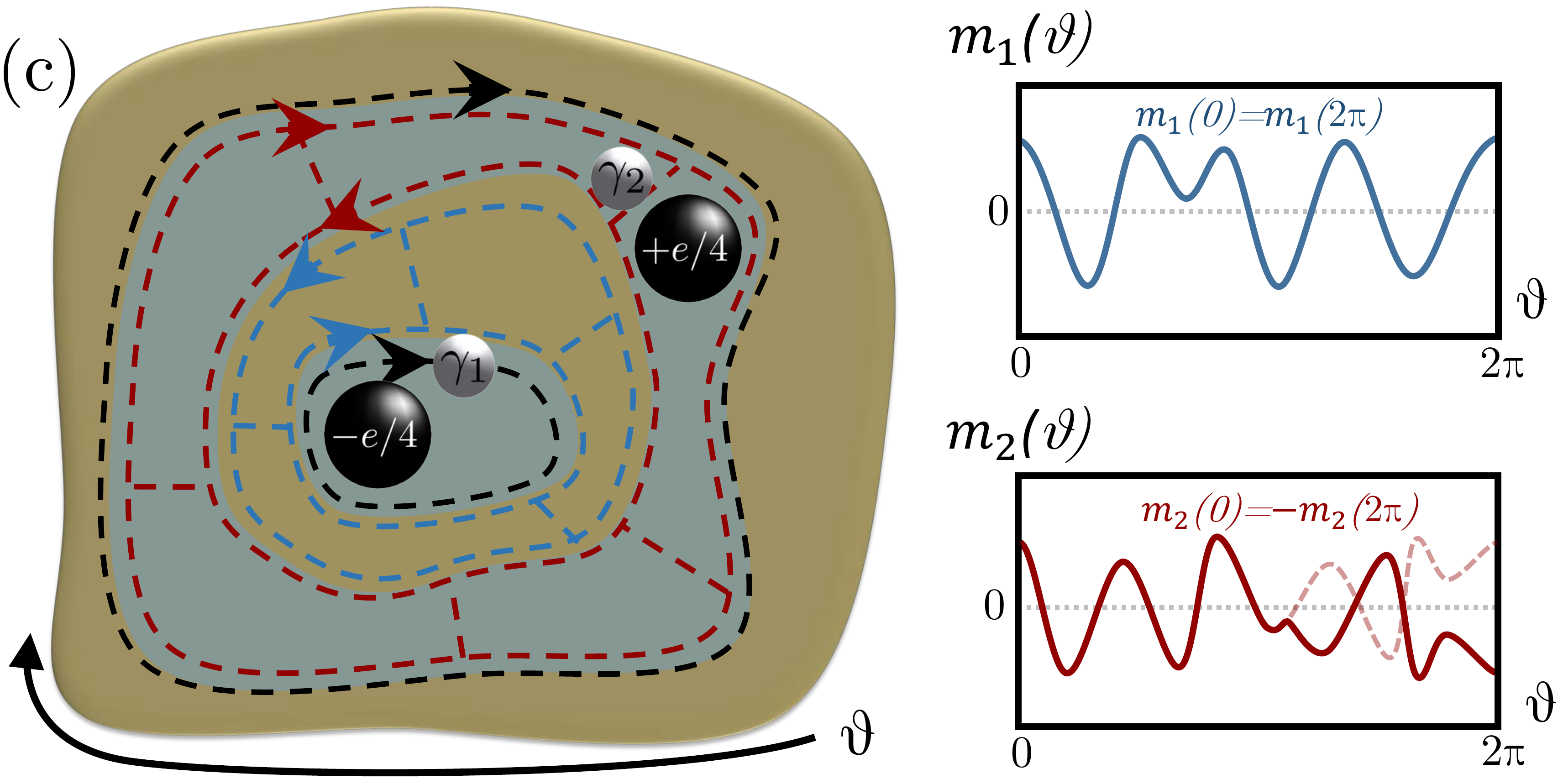}\hspace{8mm}
\includegraphics[width=.9\columnwidth]{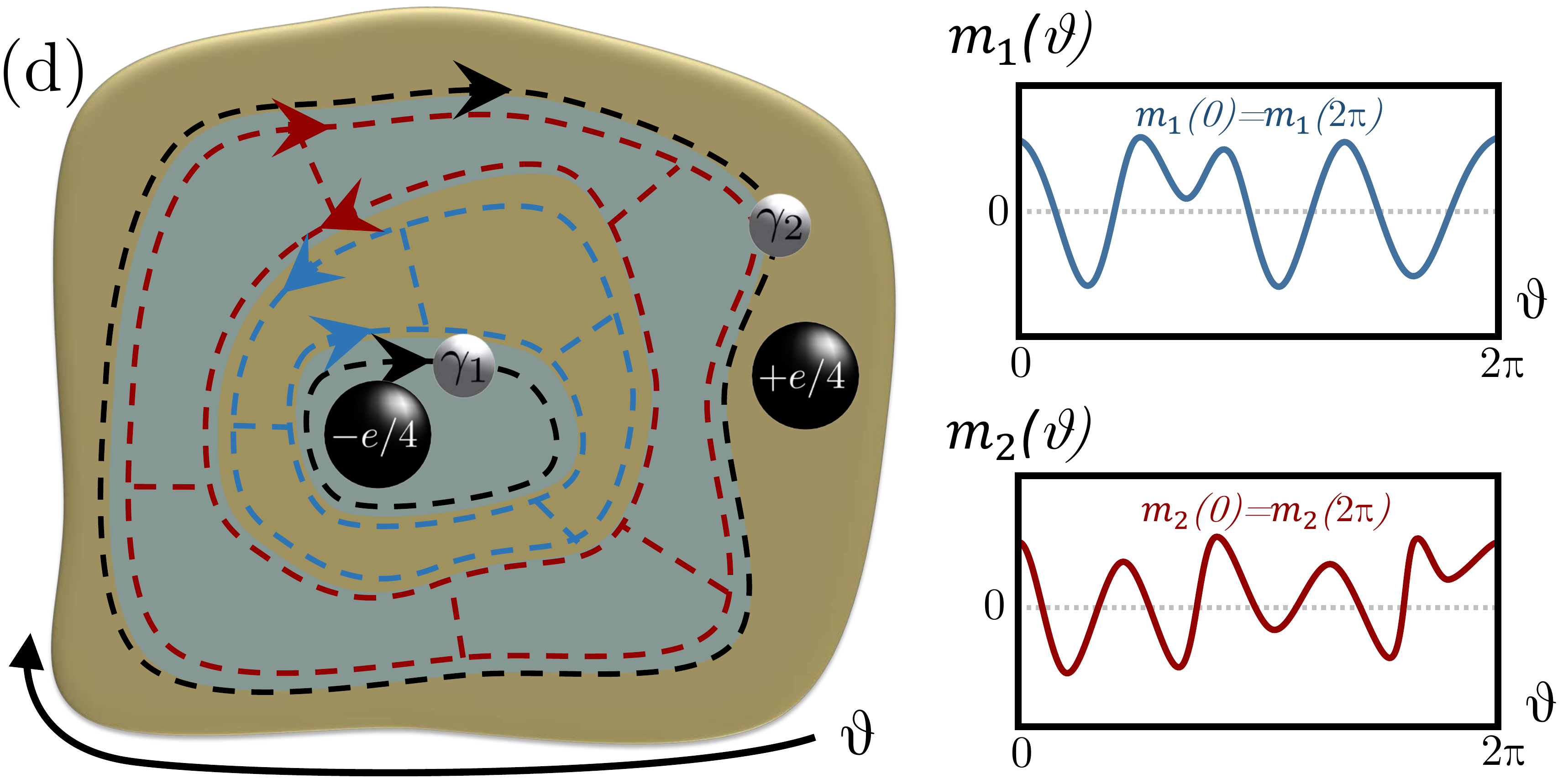}
\caption{{\bf Emergence of non-Abelian quasiparticles from puddles of Abelian phases.} For illustration, we consider concentric puddles with a pair of co-propagating Majorana fermions at each boundary. This setup is appropriate, e.g., for puddles of the $K=8$ and ($113$) states. Tunneling between neighboring edges as indicated leads to a mass term for the Majorana fermion (with spatially varying magnitude and sign). At the outermost and innermost puddle, a chiral Majorana channel remains without partner; there is consequently no mass term but finite size gaps of size $E \sim 1/L_\text{sample}$ and $E \sim 1/L_\text{puddle}$, respectively. (a) In the absence of quasiparticles, or equivalently, when the central puddle hosts a $+e/4$ quasiparticle and a $-e/4$ quasihole, the boundary conditions are such that each mass term crosses zero an \text{even} number of times upon encircling the central puddle. Consequently, there are no exact zero-energy states in the system. (b) Separation of quasiparticle and quasihole has two effects. Firstly, it changes the boundary condition of the unpaired innermost Majorana mode (denoted in black) to carry a single zero-energy state. In addition, it changes the boundary condition of the mass term $m_1$ (denoted in blue) to cross zero an odd number of times---which is associated with a second Majorana zero mode. (c) Further motion of the quasiparticle changes the boundary condition back to even for $m_1$ and turns it into odd for $m_2$. Consequently, there is no longer a zero mode associated with $m_1$, but instead there is one for $m_2$ (denoted in red). (d) Finally, when the quasiparticle reaches the outer sample boundary, the boundary conditions for all mass terms are even. However the spectrum of the unpaired outermost chiral Majorana fermion is changed to accommodate an exact zero-energy state. (Strictly speaking, there is an energy splitting between the two zero-energy states that is exponentially small in the distance between them.)}
\label{Fig:abelnonabel}

\end{figure*}

The mechanism that we propose paves the way for states composed of Abelian puddles to be macroscopically non-Abelian, and vice versa. For a state to be one of the non-Abelian states we consider, a quasiparticle should carry a Majorana zero mode, while for Abelian states there are no such zero modes. (Note that in the system we consider, there is no symmetry that protects pairs of Majorana modes from gapping out).

For concreteness, we consider first the emergence of a non-Abelian phase from Abelian puddles. When a quasiparticle/quasi-hole (\textit{qp}-\textit{qh}) pair is formed within a single puddle, they do not carry Majorana zero modes. As one of the two crosses the border to the neighboring puddle, two Majorana zero modes are created. To illustrate this process we consider a series of concentric puddles alternating between the $K=8$ and $(113)$ states, with the innermost one being, say, a $K=8$ state [see Fig.~(\ref{Fig:abelnonabel})]. Two co-propagating Majorana modes flow at the interface between every two puddles, with a direction of flow that alternates between interfaces. The non-Abelian PH-Pfaffian is created when one pair of counter-propagating Majorana modes is gapped by tunneling across a $K=8$ region (shown in red), while another pair is gapped by tunneling across a $(113)$ region (shown in blue). Each tunneling term is described in terms of a mass term $m(x)$, which may depend on position (see Fig.~\ref{Fig:abelnonabel}) . One Majorana mode is then left ungapped at the edge of the sample, and another is left ungapped around the innermost puddle (both shown in black).  Now, imagine that the \textit{qp}-\textit{qh} pair is created in the innermost puddle [Fig.~\ref{Fig:abelnonabel} (a)], and the \textit{qp} is taken away to a different puddle [Fig.~\ref{Fig:abelnonabel} (b)].  The presence of the \textit{qh} that is left behind in the innermost puddle changes the boundary conditions of the ungapped Majorana mode that encircles the puddle, and induces a zero energy state in that mode (labeled $\gamma_1$). The \textit{qp}, in turn, changes the boundary conditions on the mass term $m(x)$ that corresponds to the tunneling across the puddle on which it resides. As a consequence, this mass must cross zero, and a zero energy mode is created within that puddle, close to the \textit{qp} (labeled $\gamma_2$). The formation of the two zero energy modes attached to the \textit{qp} and \textit{qh} makes them non-Abelian.

In the inverted case, where the microscopic puddle are non-Abelian, say a Pfaffian and a PH-Pfaffian, and the macroscopic state is Abelian, the transmutation of the microscopically non-Abelian \textit{qp}s and \textit{qh}s into Abelian ones may be understood in a similar way. In this case when the \textit{qp}-\textit{qh} pair is created, they do carry intra-core zero-energy Majorana states. Those zero energy states hybridize with the states created by the mechanism we described above, leaving the \textit{qp} and \textit{qh} without zero energy states, and hence Abelian.

In the presence of disorder, where the mass term fluctuates as a function of position, we expect a closure of the energy gap by low energy localized excitations. The macroscopic nature of the \textit{qp}s and \textit{qh}s then  becomes evident only when their separation is larger than the coherence length.

\end{document}